\documentclass[conference]{IEEEtran}
\usepackage{amsmath,graphicx}
\usepackage{algorithmic}

\author{Gitanjali Bhutani\\Alcatel-Lucent Technologies India Pvt. Ltd.\\ bhutanig@alcatel-lucent.com}
\title{A Near-Optimal Scheme for TCP ACK Pacing to Maintain Throughput in Wireless Networks}
\date{}
\begin{document}
\maketitle
\begin{abstract}
  The advent of fourth generation technologies in wireless networks and the rapid growth of 3G have heralded an era that will require researchers to find reliable and easily implement-able solutions to the problem of poor TCP performance in the wireless environment. Since a large part of the Internet is TCP-based, solving this problem will be instrumental in determining if the move from wired to wireless will be seamless or not. This paper proposes a scheme that uses the base station's ability to predict the time at which the link may be going down and to estimate the period for which the mobile would be unreachable due to conditions like fading. By using cross-layer and ACK pacing algorithms, the base station prevents the fixed host from timing out while waiting for ACKs from the mobile. This in turn prevents TCP on the fixed host from bringing down the throughput drastically due to temporary network conditions, caused by mobility or the unreliability of wireless links. Experimental results indicate a reasonable increase in throughput when the ACK holding scheme is used.
\end{abstract}

\section{Introduction}
\label{sec:introduction}
With wireless devices becoming an integral part of everybody's life, there is a great emphasis on improving the offered quality of service and data rates, while increasing the range of applications that can be accessed using these devices. The third generation of wireless networks has provided a big boost to the popularity of wireless devices, with their emphasis on data and Internet access in addition to the support for voice calls. However, to enable a seamless movement from the wired to the wireless access domain, it is important to solve impending problems that reduce the achievable throughput and the efficiency of these networks. One such problem is the reduction in throughput observed due to TCP's congestion control mechanisms that come into operation during periods of mobility or fading of a wireless link.

While this problem needs to be addressed immediately, it is important to keep in mind that most of the current Internet infrastructure is TCP-based and hence, any change proposed must not entail large changes to the Internet infrastructure. Massive changes to the operation of the protocols may introduce the risk of destabilizing the entire infrastructure while also incurring large overheads in terms of effort and cost. The literature in this area has proposed several mechanisms, some requiring changes to only the mobile and some others requiring changes to the mobile as well as the base station. This paper addresses this requirement, by only requiring changes to the TCP protocol and to the link layer of the base station. The changes made to the TCP protocol are only invoked when the base station detects that the signal from the mobile is fading.
We use a proactive scheme to protect the mobile from having its throughput reduced due to a short duration of bad signal quality. It makes use of a cross-layer scheme to provide the TCP layer with updated information about the network conditions in the event of bad quality links. The ACK holding scheme discussed in this paper is a hybrid of a cross-layer mechanism coupled with a proactive scheme to prevent the wired host from taking extreme measures when the mobile experiences fading.

The ACK holding scheme proposed here, involves interaction between the link layer and TCP. The link layer uses the strength of the signal received from the mobile to determine if there is a possibility that the mobile will not be reachable in the near future. In cases where the loss of signal is due to conditions like fading, it even calculates an approximate time for which the mobile may not be reachable.  The link layer then sends the TCP layer an indication specifying that a disconnection may happen soon and the base station should start storing ACKs. The TCP layer will then spread these ACKs across the complete time interval that the mobile is estimated to be unreachable, while trying to ensure that none of these packets for which the ACKs are intended will timeout. By performing such ACK pacing, the base station ensures that during this period of temporary outage, the fixed host does not send across a large amount of data to further worsen the situation.

The paper is organized as follows: we look at related work done in the area of TCP performance and cross-layer mechanisms in Section~\ref{sec:literature-survey}. Section~\ref{sec:overview-ack-holding} contains an overview of the ACK holding scheme. Section~\ref{sec:implementation} discusses the implementation of this scheme, while Section~\ref{sec:experimental-results} gives the experimental results. Section~\ref{sec:disc-concl} concludes the discussion together with proposals for further work in this area. 

\section{Literature Survey}
\label{sec:literature-survey}
The enhancement of TCP to improve its performance in wireless networks has been a very active research area. Correspondingly, there is a large amount of literature, proposing different solutions to this problem. These solutions can be divided into three different categories, as proposed by~\cite{Balakrishnan1997}: Link-Layer schemes, End-to-End schemes and Split-Connection schemes.

Link Layer Schemes: This category of solutions involves link layer retransmissions (local retransmissions) to keep the TCP layer transparent to losses, to prevent triggering of the congestion control mechanisms. The most well known solution in this category is the Snoop protocol~\cite{Balakrishnan1995}. The Snoop protocol uses TCP duplicate acknowledgements to detect packet losses and initiate a link-level local retransmission. The base station caches every packet that has been sent to the mobile and keeps track of TCP acknowledgements to detect packet loss. By not propagating the duplicate ACK, it prevents the sender from invoking congestion control. Another scheme called delayed DupAcks~\cite{Vaidya2002} is based on Snoop~\cite{Balakrishnan1995}, but is TCP-unaware. 

End to End Solutions: End to end solutions comprise mechanisms like Selective Acknowledgements (SACK), which is an enhancement of the fast recovery scheme in TCP. In the SACK scheme, the duplicate ACK contains the sequence number of the packet that triggered the duplicate ACK. Another famous end-to-end scheme is the Explicit Loss Notification (ELN) scheme. If a packet is lost due to lossy links, a bit is set in the ACK, to indicate to the sender that congestion control mechanisms need not be invoked. Freeze-TCP~\cite{Goff2000} is another end-to-end mechanism that involves the receiver, advertising a window of zero, when it detects an impending disconnection. This causes the sender to freeze in its current state until the advertised window size is increased. 

Split-connection schemes: These schemes attempt to solve the problem by splitting the TCP connection between the fixed host and mobile host into two separate connections –- one from the fixed host to the base station and the second from the base station to the mobile host. The main aim of this technique is to keep the fixed host independent of the problems of wireless connectivity. Indirect TCP~\cite{Bakre1995} is an example of a split connection scheme. An analysis of the performance of each of these techniques leads to the conclusion that TCP-aware link layer protocols provide a better throughput as compared to the other mechanisms and the current TCP implementations~\cite{Balakrishnan1997}. 

In addition to the above, there exist other mechanisms to adapt TCP to wireless environments, including ILC Protocol~\cite{Chinta2003} and ACK-Pacing~\cite{Cho2003}. 

\section{Overview of ACK Holding Scheme}
\label{sec:overview-ack-holding}
As discussed in Section~\ref{sec:introduction}, the ACK holding scheme is a hybrid of a cross-layer scheme and a proactive method of operation. This scheme requires changes to be made to the link-layer and the TCP layer at the base station. The mobile and the fixed host remain unchanged. 

The link layer constantly monitors the strength of the signal received from the mobile, when TCP data is being exchanged between the fixed host and the mobile host. When the strength of the signal falls below a certain threshold, the link layer assumes the effects of fading or interference and concludes that the mobile may be temporarily unreachable within a certain time period. It also calculates the approximate duration for which the mobile may be unreachable. One method for doing this is proposed in~\cite{multipath}. Further discussion on the link layer mechanisms to achieve this forecasting is beyond the scope of this paper. In addition to this, the link layer must also continuously calculate the Round Trip Time (RTT) to the mobile. Once, the link layer has detected the weakening of the signal, it sends the TCP layer an indication message that specifies the TCP connection(s) corresponding to the weak link, the approximate time for which the link may be down and the mean RTT to the mobile.

Upon receipt of this message, the TCP layer knows that the link layer will now send all ACKs corresponding to this TCP connection to it and it now needs to use the ACK holding algorithm to determine the time at which to send out the next ACK for this TCP connection. In addition, after receipt of this indication, TCP is also required to cache the packets intended for this mobile. TCP needs to cache these packets to avoid packet loss and subsequent timeouts. Once, the link is further weakened and the link layer detects that any more data on this link may be lost, it sends an indication to TCP to specify that the link is now down. This indicates to TCP that it will receive no more ACKs and the ACKs currently in its cache are the ones that need to be scheduled. Keeping this behavior in mind, the ACK holding algorithm should meet these constraints:
\begin{enumerate}
\item Pace ACKs so that they are being sent over the entire estimated duration of mobile being unreachable
\item Pace ACKs so as to avoid timeouts on the fixed host for any packet for which the base station has received an ACK from the mobile
\item Pace ACKs so that RTT and retransmission timeout (RTO) on the fixed host do not increase uncontrollably
\item Use a maximum of two duplicates of each ACK, if necessary to fulfill the above requirements
\end{enumerate}
In order to meet the above requirements, the base station requires TCP on the fixed host to implement the timestamps option defined in RFC 1323~\cite{Jacobson1992}. These timestamps are needed for the base station TCP to be able to calculate the RTT from the fixed host to the base station. 

\noindent The ACK holding algorithm operates as follows:
\begin{description}
\item[Step 1] Based on 
  \begin{enumerate}
  \item Current estimate of RTT to the fixed host (R)
  \item Current estimate of RTT to the mobile (r)
  \end{enumerate}
  Calculate the approximate retransmission timeout on the fixed host for the first packet whose ACK it is holding. Let us call this t.
\item[Step 2] Based on
  \begin{enumerate}
  \item Number of ACKs present (n)
  \item Approximate duration for which the mobile is going to be unreachable (d)
  \item The estimated current retransmission timeout at the fixed host (t)
  \end{enumerate}
  Calculate the time at which the ACKs are to be sent out, while optimizing the RTO at the fixed host.
\end{description}

\noindent TCP at the base station then sends out the ACKs at the time intervals determined. Meanwhile, if there are packets arriving for the mobile, TCP uses the timestamps in these packets in order to update its estimate of RTT, RTO and eventually the time at which the ACK is to be sent to the fixed host. 

\begin{figure}
  \centering
  \includegraphics[scale=0.7]{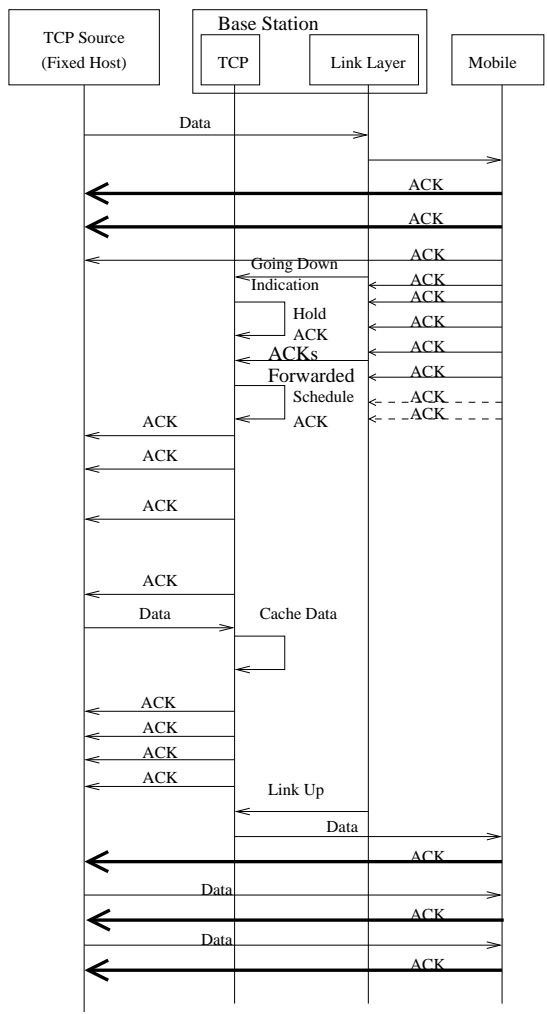}
  \caption{ACK Holding Algoritm operation}
  \label{fig:seq-diag}
\end{figure}

All incoming packets for the mobile need to be cached by the base station. The base station reserves space that is equal to the maximum advertised window size in the ACKs that it holds. When the mobile is accessible, and the link layer detects that the signal strength is above the threshold, it sends an indication to the TCP layer to stop holding ACKs and send out any remaining ACKs immediately. TCP will also transmit all cached packets to the mobile and move to the normal mode of operation. Figure~\ref{fig:seq-diag} shows the complete operation of this ACK holding scheme.

\section{Implementation}
\label{sec:implementation}
In this section, we discuss the implementation of the ACK holding scheme. For the cross layer interaction between TCP and the link-layer, the following new messages will be introduced:
\begin{enumerate}
\item Link Going Down Indication: Indicates that TCP must start holding ACKs. The contents of this message are:
  \begin{enumerate}
  \item Current RTT to mobile
  \item Approximate duration for which the link is expected to be down.
  \end{enumerate}
\item Link Gone Indication: Indicates to TCP that no more ACKs are expected. TCP can now start calculating scheduling times for the ACKs that are present. 
\item Link Up Indication: Indicates to TCP that the link is up and TCP must now send out all the ACKs to the fixed host and all the cached packets to the mobile.
\end{enumerate}

The TCP layer in the base station will be augmented with the ACK holder module which is responsible for the implementation of this scheme. The pseudo code for the working of the ACK holder algorithm is as follows:

\begin{verbatim}
Begin:
  Wait on message(m) from link layer;
  Switch (m->type) {
    Case LINK_GOING_DOWN_IND:
      modeOfOperation = HOLD_ACK;
      Spawn ACK holding thread;
    
    Case LINK_GONE_IND:
      modeOfOperation = PACE_ACK;
      Send pacing indication to 
        ACK holding thread;
		
    Case LINK_UP_IND:
      modeOfOperation = FLUSH_ACK;
      Send ACK flush indication 
        to ACK holding thread;
  }
End
\end{verbatim}
The pseudo code for the ACK holding thread is as follows:
\begin{verbatim}
Begin:
  Wait on message queue for TCP ACKs
    or indications from main thread 
    of ACK holder (a);

  Switch(a->type) {
    Case TCP_ACK:

    Case TCP_PKT:
      Store(a);
		
    Case PACE_IND:
      Determine max window size (w) 
        in ACKs; 
      Reserve space for w bytes of 
        data from fixed host;
      ScheduleAlgorithm();
		
    Case FLUSH_IND:
      Send ACKs to fixed host;
      Send cached packets to mobile;
  }
End
\end{verbatim}
The \texttt{ScheduleAlgorithm()} procedure must determine the time at which to send out each ACK based on the constraints discussed in Section~\ref{sec:overview-ack-holding}. The mechanism for doing this is explained in Section~\ref{sec:derivation}.

\section{Derivation}
\label{sec:derivation}
This section presents the derivation of the equations and the algorithm that is used to schedule the ACKs to meet the constraints described in Section~\ref{sec:implementation}. As described in~\cite{Stevens1994}, the iterative mechanism for calculating retransmission time outs by TCP is governed by the following equations:
\begin{align}
  \label{eq:rto-calculation-tcp-delta}
  \delta_i &= x_i - \mu_{i-1}\\
  \label{eq:rto-calculation-tcp-mu}
  \mu_i &= \mu_{i-1} + g \cdot \delta_i\\
  \label{eq:rto-calculation-tcp-sigma}
  \sigma_i &= \sigma_{i-1} + h \cdot \left(\left|\delta_i \right| - \sigma_{i-1}\right)\\
  \label{eq:rto-calculations-tcp-rto}
  \text{RTO}_i &= \mu_i + 4 \sigma_i
\end{align}
where $x_i$ denotes the delay round-trip time associated with packet $i$. $\mu$ and $\sigma$ are estimates of the mean and standard deviations in the round-trip times observed on the channel, as was proposed by \cite{Jacobson1988}. The work also suggested use of $g = \tfrac{1}{8}$ and $h = \tfrac{1}{4}$ for TCP implementations, as these are near optimal values for appropriate traffic shaping and also render the calculations easy as the operations involved are simple right shift bit operations. 
 
In order to prevent the fixed host from invoking its congestion control procedures when the mobile goes out of the service area, the proposed scheme divides the accumulated ACKs into two subsets. It utilizes the first subset of ACKs for increasing the RTO calculations at the fixed host, so that each ACK can be used to make the source wait longer. After having covered a large portion of the total time with the first subset of ACKs, the second subset of ACKs are then sent back to the source at a constant and smaller rate. This ensures that the calculated value of RTO that was made to increase by use of ACKs in the first subset, is returned to smaller manageable numbers by the time the mobile comes back into service.  

We denote the total number of ACKs stored by the base station as $N$. The inter-ACK arrival time, as observed by the TCP source, for the $i^\text{th}$ ACK is denoted by $x_i$. $T$ denotes the amount of time after which the mobile is estimated to come back up and $n$ denotes the number of ACKs made to wait as long as possible right at the start of the scheme. Note that we use the next set of $(N-n)$ packets to bring the RTO (and consequently $\mu$ and $\sigma$) values within acceptable bounds for a normal connection to operate properly when the mobile comes back up at the estimated time. This is done by sending these $(N-n)$ ACKs with a small fixed amount of time between them, denoted by $\theta$. 

\begin{figure}
  \centering
  \includegraphics[width=0.8\columnwidth]{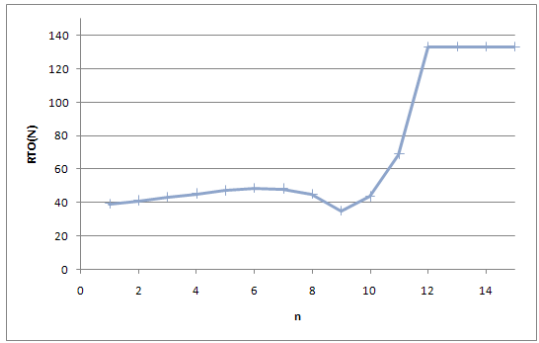}
  \caption{$\text{RTO}_N$ as a function of $n$ for $T = 1000, N = 30, \mu_0 = 1, \sigma_0 = 0.3$}
  \label{fig:rto-function-n}
\end{figure}

As the first $n$ ACKs are made to wait as long as possible without causing a retransmission from the TCP source, we set $x_i = \text{RTO}_{i-1}; \forall i\leq n$. Hence $\text{RTO}_0 = \mu_0 + 4 \sigma_0 = x_1$, $\mu_1 = \mu_0 + \frac{\sigma_0}{2}$ and $\sigma_1 = \frac{7}{4}\sigma_0$. Similarly, $\text{RTO}_1 = \mu_0 + \frac{15}{2} \sigma_0 = x_2$, $\mu_2 = \mu_0 + \frac{11}{8}\sigma_0$ and $\sigma_2 = \left(\frac{7}{4}\right)^2 \sigma_0$, thereby leading to $\text{RTO}_2 = \mu_0 + \frac{109}{8} \sigma_0$. Expanding similarly, we find that for $i \leq n$, $\sigma_i = \left(\frac{7}{4}\right)^i \sigma_0$ and $\mu_i = \mu_0 + \left(\frac{\sigma_{i-1}}{2} + \frac{\sigma_{i-2}}{2} + \frac{\sigma_{i-3}}{2} + \cdots\right)$. Therefore, the values of $\sigma$, $\mu$ and RTO after the first $n$ ACKs have been sent out (denoted as $\sigma_n$, $\mu_n$ and $\text{RTO}_n$ respectively) can be determined as in Equations~(\ref{eq:sigma-n}), (\ref{eq:mu-n}) and (\ref{eq:RTO-n}). 
\begin{equation}
  \label{eq:sigma-n}
  \sigma_n = \left(\frac{7}{4}\right)^n \sigma_0
\end{equation}

\begin{equation}
  \label{eq:mu-n}
  \mu_n = \mu_0 + \frac{2}{3} \sigma_0 \left[ \left(\frac{7}{4}\right)^n - 1\right]
\end{equation}

\begin{equation}
  \label{eq:RTO-n}
  \text{RTO}_n = \mu_0 + \sigma_0 \left[ \frac{14}{3} \left(\frac{7}{4}\right)^n - \frac{2}{3}\right]
\end{equation}
We denote by $S(n)$ the amount of time the TCP source has been made to wait using the first $n$ ACKs alone. Note that this would equal the sum of inter-ACK times for the first $n$ packets, thereby making $S(n) = \sum_{i=1}^n x_i$. The value of $S(n)$ is calculated in Equation~(\ref{eq:sum-RTO-n}).
\begin{equation}
  \label{eq:sum-RTO-n}
  S(n) =  \sum_{i=0}^{n-1} \text{RTO}_i = n \mu_0 + \sigma_0 \left\{ \frac{56}{9} \left[ \left(\frac{7}{4}\right)^n - 1\right] - \frac{2n}{3}\right\} 
\end{equation}
To determine the final values of $\sigma$, $\mu$ and RTO after all $N$ ACKs have reached the TCP source, denoted by $\sigma_N(n)$, $\mu_N(n)$ and $\text{RTO}_N(n)$ respectively, we set $x_i = \frac{(T-S(n))}{(N-n)} \equiv \theta; \forall i \in \{n+1, \ldots, N\}$. Note that all ACKs from $(n+1)^\text{th}$ ACK till the $N^\text{th}$ ACK are spaced out equally, with all of their inter-ACK arrival times being set equal to $\theta$. We find that $\delta_{n+1} = (\theta - \mu_n)$, $\mu_{n+1} = \frac{7}{8}\mu_n + \frac{1}{8}\theta$ and $\sigma_{n+1} = \frac{3}{4}\sigma_n + \frac{1}{4} \cdot |\theta - \mu_n|$. Similarly, $\delta_{n+2} = \frac{7}{8}(\theta - \mu_n)$, $\mu_{n+2} = \left(\frac{7}{8}\right)^2(\mu_n - \theta) + \theta$ and $\sigma_{n+2} = \left(\frac{3}{4}\right)^2 \sigma_n + \frac{1}{4} \left(\frac{3}{4} + \frac{7}{8}\right) \cdot |\theta - \mu_n|$. Expanding similary, we find that $\delta_{n+i} = \left(\frac{7}{8}\right)^{i-1} (\theta - \mu_n)$, $\mu_{n+i} = \left(\frac{7}{8}\right)^i (\mu_n - \theta) + \theta$ and $\sigma_{n+i} = \left(\frac{3}{4}\right)^i + 2 |\theta - \mu_n| \cdot \left[\left(\frac{7}{8}\right)^i - \left(\frac{3}{4}\right)^i\right]$. Hence expressions for $\mu_N(n)$ and $\sigma_N(n)$ can be written as
\begin{equation}
  \label{eq:mu-N}
  \mu_N(n) = \left(\frac{7}{8}\right)^{(N-n)} \mu_n + \left[ 1 - \left( \frac{7}{8}\right)^{(N-n)}\right] \theta
\end{equation}

\begin{multline}
  \label{eq:sigma-N}
  \sigma_N(n) = \left(\frac{3}{4}\right)^{(N-n)} \sigma_n \\+ 2 \left[\left(\frac{7}{8}\right)^{(N-n)} - \left(\frac{3}{4}\right)^{(N-n)}\right] |\mu_n - \theta|
\end{multline}
$\text{RTO}_N$ can now be calculated as per Equation~(\ref{eq:rto-calculations-tcp-rto}) as
\begin{multline}
  \text{RTO}_N(n) = \left(\frac{7}{8}\right)^{(N-n)} \mu_n + 4\left(\frac{3}{4}\right)^{(N-n)} \sigma_n 
  \\+ \left[1-\left(\frac{7}{8}\right)^{(N-n)}\right] \theta(n)  
  \\+ 8\left[\left(\frac{7}{8}\right)^{(N-n)} - \left(\frac{3}{4}\right)^{(N-n)}\right] |\mu_n-\theta(n)|
  \label{eq:RTO-N}
\end{multline}
However, following the traditional method of setting the partial derivative of $\text{RTO}_N(n)$ with respect to $n$ to zero for finding the minima in Equation~(\ref{eq:RTO-N}) does not yield a closed form solution. For given values of $T, N, \mu_0 \text{ and } \sigma_0$, the typical variation in $\text{RTO}_N$ as $n$ is varied is shown in Figure~\ref{fig:rto-function-n}. Note that while the function derivative of $\text{RTO}_N$ appears to have multiple solutions, it has only one minima. Hence, the value of $n$ that minimizes $\text{RTO}_N$ can be calculated using a single step forward algorithm shown below. 

\begin{verbatim}
Initialize: 
  Set RTO_prev := RTO_curr := RTO_N(0);
  Set RTO_next := RTO_N(1);
  Iterate Over N:
    If ((RTO_curr <= RTO_prev) and 
        (RTO_curr <= RTO_next))
      Stop;
    Else
      Increment n;
      Set RTO_prev := RTO_curr;
      Set RTO_curr := RTO_next;
      Set RTO_next := RTO_N(n+1);
    EndIf
  EndIterate
End
\end{verbatim}

\section{Experimental Results}
\label{sec:experimental-results}
\begin{figure}
  \centering
  \includegraphics[width=0.8\columnwidth]{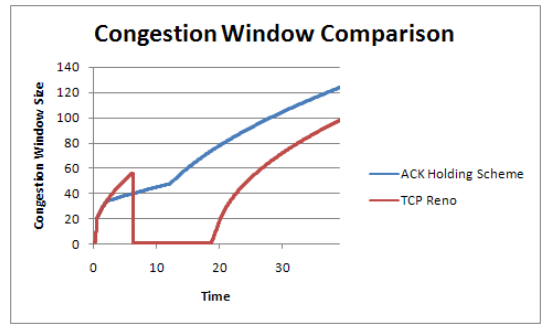}
  \caption{Congestion window comparison for single interval of fading}
  \label{fig:results-time}
\end{figure}

The ACK holding scheme was implemented in ns-2 and its throughput compared to that of TCP without the ACK holding scheme, in the presence of lossy links. Figure~\ref{fig:results-time} compares the congestion window size of TCP-Reno with that of TCP augmented with the ACK holding scheme, when the wireless link was down for 10 seconds during a 40 second data transfer interval. 
\begin{table}
  \centering
  \begin{tabular}{|l|c|c|c|}
    \hline
    \textbf{Disconnection Intervals} & \textbf{ACK Holding} & \textbf{TCP Reno} & \textbf{Throughput}\\
    & \textbf{Scheme} & & \textbf{Improvement(\%)}\\
    \hline
    Single intervals of 10s & 7688 & 6361 & 120.9\%\\
    Two intervals of 10s each & 3981 & 503 & 791.5\%\\
    \hline
  \end{tabular}
  \caption{Comparison of Throughput}
  \label{tab:comparison-throughput}
\end{table}
\begin{figure}
  \centering
  \includegraphics[width=0.8\columnwidth]{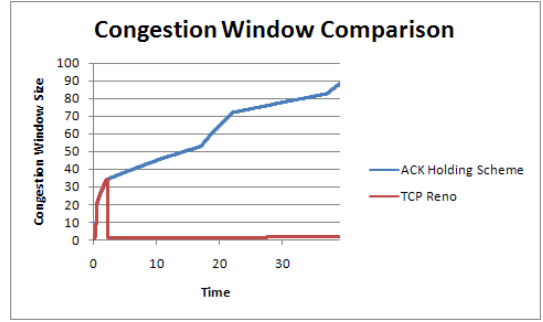}
  \caption{Congestion window comparison when fading takes place for two successive time intervals}
  \label{fig:2Interval}
\end{figure}
This figure clearly shows the positive effect of TCP's ACK holding scheme as it prevents any timeouts and hence, the reduction of the congestion window to 1. Figure~\ref{fig:2Interval} compares the congestion window size of TCP-Reno and the ACK holding scheme when there are two 10 second disconnections during a 40 second data transfer. This figure proves that the reason the ACK holding scheme performs better than the current TCP implementations is that the TCP at the sender requires no time to recover after the disconnection period. 

Table~\ref{tab:comparison-throughput} shows the throughput comparison of TCP-Reno and the ACK holding scheme for the above two conditions. It can be seen, that the ACK holding scheme, achieves throughput over 7.5 times that of the current TCP. This is due to the fact that after slow start has been invoked TCP takes some amount of time to recover and bring its data rate to what it was before the link went down. However, in the case of the ACK holding scheme, there is no such recovery needed. After the wireless link is restored, the only component that needs to recover in the presence of the ACK holding scheme is the retransmission timeout estimate. Since, the ACKs are delayed by specific intervals, the RTT estimate at the sender is disturbed, making the RTO larger. 
\begin{figure}
  \centering
  \includegraphics[width=0.8\columnwidth]{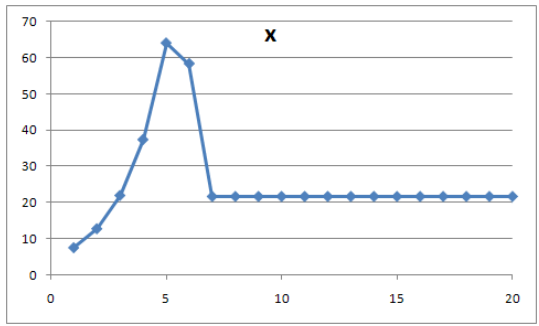}
  \caption{Optimal $x$ for a wait of 500 seconds}
  \label{fig:optimal-x-500-wait}
\end{figure}

Figure~\ref{fig:optimal-x-500-wait} shows the change in the inter-ACK time using the equations in Section~\ref{sec:derivation}, when the base station has 20 ACKs to pace over a period of 500 seconds. The inter-ACK time initially increases for the first 5 ACKs. The inter-ACK time then decreases until it eventually stabilizes to a small constant value that will be instrumental in reducing the final RTO after the ACK holding algorithm terminates.

\begin{figure}
  \centering
  \includegraphics[width=0.8\columnwidth]{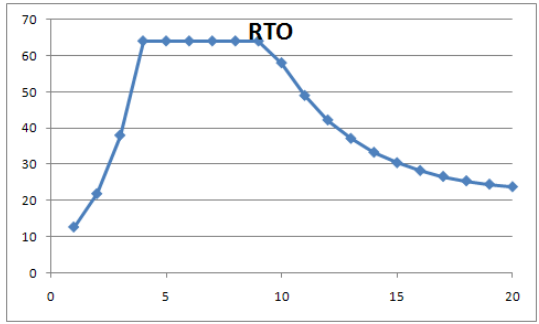}
  \caption{Variation of RTO with optimal $x$ for a wait of 500 seconds}
  \label{fig:rto-variation-500-wait}
\end{figure}
\begin{figure}
  \centering
  \includegraphics[width=0.8\columnwidth]{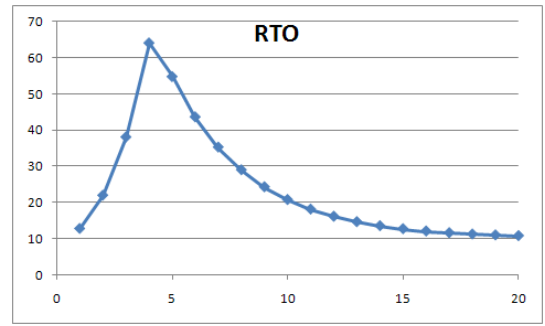}
  \caption{Variation of RTO with optimal $x$ for a wait of 250 seconds}
  \label{fig:rto-variation-250-wait}
\end{figure}

Figure~\ref{fig:rto-variation-500-wait} shows the variation in RTO value for the above case. In keeping with the increase in inter-ACK time, the RTO value calculated at the source first increases and then stabilizes at a higher value before decreasing and stabilizing at a lower value.
\begin{figure}
  \centering
  \includegraphics[width=0.8\columnwidth]{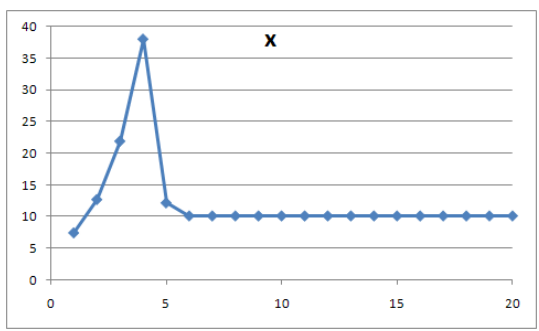}
  \caption{Optimal $x$ for a wait of 250 seconds}
  \label{fig:optimal-x-250-wait}
\end{figure}
Figure~\ref{fig:optimal-x-250-wait} shows the inter-ACK time variation using the ScheduleAlgorithm of Section~\ref{sec:implementation} to a set of 20 ACKs paced over a period of 250 seconds. Figure~\ref{fig:rto-variation-250-wait} shows the RTO variation for the same case.

\section{Discussion and Conclusions}
\label{sec:disc-concl}
This paper proposed a proactive scheme that used cross-layer techniques at the base station to inform TCP of a temporary loss of connectivity to a mobile. The base station TCP then used the estimate of the link down period to schedule the ACKs received from the mobile in such a way that the fixed host was kept transparent to the loss of connectivity. In Section~\ref{sec:experimental-results}, we saw that this improves the throughput by almost 7.5 times as compared to the current TCP implementation. This proves that the ACK holding schemes provides large improvements in scenarios where there are multiple disconnections in a single data transfer session. 


In Section~\ref{sec:derivation}, we have assumed that accurate predictions are available of duration for which the mobile would remain disconnected. It may therefore seem that the scheme is dependent on availability of such an accurate predictor. However, note from Figures~\ref{fig:rto-variation-500-wait} and~\ref{fig:rto-variation-250-wait} that the final values of RTO do not vary considerably even for large changes in values of $T$. Hence, in the absence of an accurate prediction of $T$, we suggest that this scheme should be implemented with over-estimates of $T$ rather than under-estimates.

The results were obtained by modifying ns-2 to hold and schedule ACKs in order to simulate the ACK holding scheme and drop ACKs and packets to simulate the working of normal TCP in the event of a disconnection. Using the results obtained by deploying this scheme in a real network, will allow further fine tuning of the algorithm to arrive at an optimal solution. Observing the effects that this scheme has in a real network will be the most important test for it. 

\bibliographystyle{IEEEtran}
\bibliography{network}

\end{document}